\documentclass[11pt]{article}

\newsavebox{\foobox}
\newcommand{\slantbox}[2][0]{\mbox{%
        \sbox{\foobox}{#2}%
        \hskip\wd\foobox
        \pdfsave
        \pdfsetmatrix{1 0 #1 1}%
        \llap{\usebox{\foobox}}%
        \pdfrestore
}}
\newcommand\unslant[2][-.25]{\slantbox[#1]{$#2$}}

\newcommand{\mpi}{\text{\unslant[-.18]\pi}}
\newcommand{\mdelta}{\text{\unslant[-.18]\delta}}

\usepackage[left=2cm, right=2cm, top=2.5cm, bottom=2.5cm]{geometry}
\usepackage[x11names]{xcolor}
\usepackage{fancyhdr, amssymb, cancel, amsmath, graphicx, pgfplots, tikz}
\usepackage{isomath}

\usetikzlibrary{shadows}

\newcommand{\stylecolor}{blue}

\usepackage[labelfont={bf,sf, color=\stylecolor}, margin={1.5cm,0cm}]{caption}

\usepackage[colorlinks=true, urlcolor=\stylecolor!70!white, linkcolor=\stylecolor, citecolor=\stylecolor!70!white, hyperindex=true, linktocpage=true]{hyperref}

\usepackage[explicit]{titlesec}

\newcommand*\sectionlabel{}
\titleformat{\section}
  {\gdef\sectionlabel{}
   \Large\bfseries\scshape}
  {\gdef\sectionlabel{\thesection }}{0pt}
  {\begin{tikzpicture}[remember picture,overlay]
       \end{tikzpicture}
  }
\titlespacing*{\section}{0pt}{0pt}{0pt}

\newcommand*\subsectionlabel{}
\titleformat{\subsection}
  {\gdef\subsectionlabel{}
   \large\bfseries\scshape}
  {\gdef\subsectionlabel{\thesubsection  }}{0pt}
  {\begin{tikzpicture}[remember picture]
    	\draw (-0.15, 0) node[left] {\color{\stylecolor} \textsf{\subsectionlabel}};
	\draw (0.15, 0) node[right] {\color{\stylecolor} \textsf{#1}};
	\fill[color=\stylecolor] (-0.05, -0.23) rectangle (0.05, 0.23);
       \end{tikzpicture}
  }
\titlespacing*{\subsection}{-4pt}{10pt}{0pt}

\newcommand*\subsubsectionlabel{}
\titleformat{\subsubsection}
  {\gdef\subsubsectionlabel{}
   \bfseries\scshape}
  {\gdef\subsubsectionlabel{\thesubsubsection.\ \  }}{0pt}
  {\begin{tikzpicture}[remember picture]
    	\draw (0, 0) node[left] {\color{\stylecolor} \textsf{\subsubsectionlabel}};
	\draw (0, 0) node[right] {\color{\stylecolor} \textsf{#1}};
       \end{tikzpicture}
  }
\titlespacing*{\subsubsection}{-4pt}{7pt}{0pt}

\pgfplotsset{every axis legend/.append style={at={(1.02,1)},anchor=north west}}

\begin{document}

\allowdisplaybreaks

\pagestyle{fancy}
\renewcommand{\headrulewidth}{0pt}
\fancyhead{}

\fancyfoot{}
\fancyfoot[C] {\textsf{\textbf{\thepage}}}

\begin{equation*}
\begin{tikzpicture}
\draw (\textwidth, 0) node[text width = \textwidth, right] {\color{white} easter egg};
\end{tikzpicture}
\end{equation*}

\begin{equation*}
\begin{tikzpicture}
\draw (0.5\textwidth, -3) node[text width = \textwidth] {\huge  \textsf{\textbf{Operator size at finite temperature  and Planckian  \\ \vspace{0.07in}  bounds on quantum  dynamics }} };
\end{tikzpicture}
\end{equation*}
\begin{equation*}
\begin{tikzpicture}
\draw (0.5\textwidth, 0.1) node[text width=\textwidth] {\large \color{black}  \textsf{Andrew Lucas}};
\draw (0.5\textwidth, -0.5) node[text width=\textwidth] {\small\textsf{Department of Physics, Stanford University, Stanford, CA 94305, USA}};
\end{tikzpicture}
\end{equation*}
\begin{equation*}
\begin{tikzpicture}
\draw (0, -13.1) node[right, text width=0.5\paperwidth] {\texttt{ajlucas@stanford.edu}};
\draw (\textwidth, -13.1) node[left] {\textsf{\today}};
\end{tikzpicture}
\end{equation*}
\begin{equation*}
\begin{tikzpicture}
\draw[very thick, color=\stylecolor] (0.0\textwidth, -5.75) -- (0.99\textwidth, -5.75);
\draw (0.12\textwidth, -6.25) node[left] {\color{\stylecolor}  \textsf{\textbf{Abstract:}}};
\draw (0.53\textwidth, -6) node[below, text width=0.8\textwidth, text justified] {\small  It has long been believed that dissipative time scales $\tau$ obey a ``Planckian" bound  $\tau \gtrsim \frac{\hbar}{k_{\mathrm{B}}T}$ in strongly coupled quantum systems.   Despite much circumstantial evidence, however, there is no known $\tau$ for which this bound is universal.   Here we define operator size at finite temperature, and conjecture such a $\tau$:  the time scale over which small operators become large.   All known many-body theories are consistent with this conjecture.   This proposed bound explains why previously conjectured Planckian bounds do not always apply to weakly coupled theories, and how Planckian time scales can be relevant to both transport and chaos.};
\end{tikzpicture}
\end{equation*}

\tableofcontents

\begin{equation*}
\begin{tikzpicture}
\draw[very thick, color=\stylecolor] (0.0\textwidth, -5.75) -- (0.99\textwidth, -5.75);
\end{tikzpicture}
\end{equation*}

\titleformat{\section}
  {\gdef\sectionlabel{}
   \Large\bfseries\scshape}
  {\gdef\sectionlabel{\thesection }}{0pt}
  {\begin{tikzpicture}[remember picture]
	\draw (0.2, 0) node[right] {\color{\stylecolor} \textsf{#1}};
	\draw (0.0, 0) node[left, fill=\stylecolor,minimum height=0.27in, minimum width=0.27in] {\color{white} \textsf{\sectionlabel}};
       \end{tikzpicture}
  }
\titlespacing*{\section}{0pt}{20pt}{5pt}

\section{Introduction}
It has long been believed that the Planckian time scale $\frac{\hbar}{k_{\mathrm{B}}T}$ bounds dissipation and thermalization in a many-body quantum system with $N \gg 1$ degrees of freedom and few-body interactions:  \begin{equation}
\tau \gtrsim \frac{\hbar}{k_{\mathrm{B}}T}.  \label{eq:main}
\end{equation}   Driven by both theory and experiment, bounds similar to (\ref{eq:main}) have been conjectured for many physically measurable quantities.   (\emph{i}) In strongly coupled systems, including quantum critical models \cite{sachdev98, sachdev} and holographic models \cite{hubeny, festuccia,hod}, the correlation functions of spatially local operators $\mathcal{O}$ obey $\langle \mathcal{O}(t) \mathcal{O}\rangle \sim \mathrm{e}^{-\gamma t}$, where $\gamma \sim k_{\mathrm{B}}T/\hbar$.  One conjecture is  that $\gamma^{-1}$ obeys (\ref{eq:main}).   (\emph{ii}) Many experiments do not measure decay rates $\gamma$ directly, but instead measure dissipative transport coefficients such as electrical condcutivity $\sigma$.  If we write  $\sigma = \chi_{JJ} \tau_{\mathrm{tr}}$, with $\chi_{JJ}$ a suitable thermodynamic coefficient, then $\tau_{\mathrm{tr}}$ is a ``transport time".  Remarkably, the conductivity of many strongly interacting metals is consistent with $\tau_{\mathrm{tr}} \sim \frac{\hbar}{k_{\mathrm{B}}T}$ \cite{cooper, mackenzie2013, legros}, which suggests that (sometimes) $\tau_{\mathrm{tr}} \gtrsim  \frac{\hbar}{k_{\mathrm{B}}T}$ \cite{zaanen}.   (\emph{iii})  Cold atomic gases \cite{cao} and quark-gluon plasma \cite{shuryak, romatschke} have shear viscosities compatible with a Planckian bound on the viscosity $\eta$ of quantum fluids,  $\frac{\eta}{Ts} \gtrsim  \frac{\hbar}{k_{\mathrm{B}}T}$, with $s$ the entropy density \cite{kss}.  (\emph{iv})  These resistivity and viscosity bounds have been proposed to be examples of a generic Planckian diffusion bound $D\gtrsim v^2  \frac{\hbar}{k_{\mathrm{B}}T}$ \cite{hartnoll1};  experimental evidence for Planckian thermal diffusion is presented in \cite{levenson, aharon18}.     (\emph{v})  In many interacting systems, the quantum Lyapunov time obeys \cite{stanfordbound} \begin{equation}
\tau_{\mathrm{L}} \ge \frac{\hbar}{2\mpi k_{\mathrm{B}}T} . \label{eq:mss}
\end{equation}
In some theories, (\ref{eq:mss}) appears related to the proposed Planckian bound on diffusion \cite{blakeB1, blakeB2, aleiner, gu, patel, blakeB3, gu2, mendl, gouteraux, patel2, kkim1, blakeB4, matteo, erez, scopelliti, blake18, scopelliti2}.

An intuitive argument for (\ref{eq:main}) follows from the Heisenberg energy-time uncertainty principle \begin{equation}
\mathrm{\Delta}E\mathrm{\Delta}t \gtrsim \hbar, \label{eq:DeltaE}
\end{equation}
with $\tau= \mathrm{\Delta } t$ and $k_{\mathrm{B}}T=\mathrm{\Delta}E$ set by thermal fluctuations.   Unfortunately, $\mathrm{\Delta}t$ is formally a dephasing time for the many-body wave function \cite{frohlich}:  it is neither measurable nor significant when $N\gg 1$.  Proving (\ref{eq:main}), and even defining $\tau$, has remained an open problem.

None of the bounds above apply to all many-body quantum systems with few-body interactions.  (\emph{i}) Disorder gives many measurable operators a finite decay rate $\gamma$ at zero temperature in a non-interacting Fermi gas \cite{palee}.  (\emph{ii})  Planckian lower bounds on the conductivity are  violated by the residual (temperature-independent) conductivity of metals at $T=0$ \cite{palee}, and at a continuous disorder-driven metal-insulator transition.\footnote{This transition was also a counterexample to the earlier Mott-Ioffe-Regel bound \cite{ioffe,mott}, where $\tau_{\mathrm{tr}}$ was conjectured to be at least as large as the quasiparticle coherence time.}    (\emph{iii}) Bounds on $\frac{\eta}{s}$ are violated with higher derivative holographic models \cite{yaida07} and/or with anisotropy \cite{trivedi, link}; holographic anisotropic violations are resolved  in \cite{blakeB1}.  (\emph{iv}) Rigorous bounds on diffusion are upper bounds \cite{julia, gu2017chain, hartnoll1706, lucas1710, han}.    (\emph{v}) The bound (\ref{eq:mss}) does not apply to certain integrable models, including free fermions \cite{stanfordbound}.     

Is there a time scale $\tau$ for which  (\ref{eq:main}) holds in \emph{all} many-body systems?  If so, \emph{how} is (\ref{eq:main}) consistent with the systems above? Are conjectures (\emph{i})-(\emph{v}) limiting cases of (\ref{eq:main}), despite the very different correlation functions they relate to?    Why is most evidence for (\ref{eq:main}) from strongly coupled systems?   

This letter proposes a simple answer to all of these questions.  We conjecture that (\ref{eq:main}) holds in generic many-body systems with few-body interactions, so long as $\tau$ is the time after which a ``small" operator decays into a (sum of) ``large" operators.   To do so, we introduce and motivate a particular notion of operator size at finite temperature.   Although bounds (\emph{i})-(\emph{v}) are on qualitatively different correlation functions, each directly relates to our conjecture.  We further explain how quantum systems that avoid bounds (\emph{i})-(\emph{v}) are consistent with our conjecture.   Our bound is never saturated in weakly coupled systems, and is consistent with existing results in all known strongly coupled systems, including lattice models, quantum critical theories, and holographic theories.    Hence, we obtain a compelling framework where (\ref{eq:main}) appears to be a fundamental constraint on quantum dynamics.   We set $\hbar=k_{\mathrm{B}}=1$ below.

\section{Operator Size}
First, we precisely define terms introduced above.    We study many-body quantum systems with $N\gg 1$ quantum degrees of freedom (DOF).  For simplicity, the Hilbert space $\mathcal{H} = \bigotimes_{i=1}^N \mathcal{H}_i$, with $\dim(\mathcal{H}_i)=q<\infty$ for all $i$; each $i$ denotes a DOF.     Having few-body interactions means that the many-body Hamiltonian can be written as a sum of operators which act on at most $k\propto N^0$ DOF:\footnote{We allow the same set to show up multiple times in the sum below.}
\begin{equation}
H = \sum_{X=\lbrace i_1,\ldots, i_\ell\rbrace:  |X| =m\le k} \mathcal{O}_{i_1}\otimes\cdots\otimes\mathcal{O}_{i_{m}}. \label{eq:HS}
\end{equation}
Operators of this form are called $k$-local.    Here and below, $\mathcal{O}_i$ denotes an operator which acts non-trivially only on DOF $i$.   Interactions in $H$ can be ``non-local" in the conventional physics sense: DOF $i$ and $j$ may be coupled together for all pairs $i$ and $j$; $k$-locality means that $1,\ldots,N$ are not \emph{simultaneously} coupled.   For generic $H$, $k$-locality holds in a unique basis (up to local rotations) \cite{cotler}: thus, the tensor product decomposition of $\mathcal{H}$ is not arbitrary.    

Defining operator size takes more work.   Intuitively, operator size should be defined so that $k$-local operators are small, and $N$-local operators are large.    Operator time evolution $\mathcal{O}(t) = \mathrm{e}^{\mathrm{i}Ht} \mathcal{O} \mathrm{e}^{-\mathrm{i}Ht} = \mathcal{O}+ \mathrm{i}t[H,\mathcal{O}] -\frac{t^2}{2}[H,[H,\mathcal{O}]]$ makes small operators grow larger:  each $[H,\circ]$ can make a $q$-local operator $(q+k-1)$-local.    In general, for any $t\ne 0$, $\mathcal{O}(t)$ is $N$-local.  Worse, $\mathcal{O}(t)$ does not depend on the state of the quantum system: how does operator size depend on temperature?

We will address these questions one at a time.   First, a natural way to introduce temperature into operator dynamics is to define a thermal inner product on operators themselves.  After all, the time evolution of operators is linear and analogous to the Schr\"odinger equation for wave functions: denoting operator $\mathcal{O}$ as $|\mathcal{O})$, time translations are generated by $\mathcal{L}|\mathcal{O}) = |\mathrm{i}[H,\mathcal{O}])$, and $|\mathcal{O}(t)) = \mathrm{e}^{\mathcal{L}t} |\mathcal{O})$.   Our thermal inner product on the space of operators will intuitively project out operators acting on eigenstates whose energies $E$ are too large.   More precisely, let $\beta = T^{-1}$ and $\langle A\rangle = \frac{\mathrm{tr}(\mathrm{e}^{-\beta H}A)}{\mathrm{tr}(\mathrm{e}^{-\beta H})}$ denotes thermal expectation values.  One thermal inner product, which has some nice properties, is  \cite{lucasbook}
\begin{equation}
(A|B) = T\int\limits_0^\beta \mathrm{d}\lambda \; \left\langle A^\dagger  B(\mathrm{i}\lambda)  \right\rangle. \label{eq:innerproduct}
\end{equation} 
Choosing operators $A$ and $B$ with no thermal average ($\langle A\rangle = \langle B\rangle = 0$), $(A|B) = T\chi_{AB}$, with $\chi_{AB}$ the thermodynamic susceptibility.   Operators are orthogonal when their thermodynamic fluctuations decouple.  Furthermore, $\mathcal{L}$ is antisymmetric:  $(A|\mathcal{L}|B)  = -(B|\mathcal{L}|A)$, so a Hermitian operator $|\mathcal{O}(t))$ has constant length under time evolution.    In summary, there is a natural temperature dependent ``metric" on operator space.

To answer the second question, we define size not by the $q$-locality of operators, but by a matrix $\mathcal{S}$, acting on the vector space of operators equipped with inner product (\ref{eq:innerproduct}).  Recalling that each quantum DOF has $q$ levels, define traceless Hermitian operators $T^a_i$ ($a=1,\ldots,q^2-1$) for DOF $i$.   Choosing normalization $\mathrm{tr}(T^a T^b) = q\mdelta^{ab}$, the matrix elements of $\mathcal{S}$ are
\begin{equation}
(A|\mathcal{S}|B) = \sum_{i=1}^N \sum_{c=1}^{q^2-1} \frac{([A,T^c_i]|[B,T^c_i])}{2q^2}.  \label{eq:TinfOTOC}
\end{equation}
One justification for this definition comes from considering the infinite temperature limit of (\ref{eq:innerproduct}):  $(A|B) = q^{-N}\mathrm{tr}(A^\dagger B)$.    We first fix $i$ and sum over $c$ in (\ref{eq:TinfOTOC}).  Let $A=A_{-i}\otimes T^a_i$ and $B=B_{-i}\otimes T^b_i$, with $A_{-i}$ and $B_{-i}$ acting on all DOF except $i$.    Using $\mathrm{SU}(q)$ identities for $[T^a_i, T^b_i] = \mathrm{i}f^{abc}T^c_i$ (repeated $abc$ indices summed over) we obtain \begin{equation}
\frac{([A,T^c_i]|[B,T^c_i])}{2q^2} =   \frac{\mathrm{tr}(A_{-i}B_{-i})}{q^{N-1}}  \frac{f^{acd}f^{bce}}{2q^2} \frac{\mathrm{tr}(T^d_i T^e_i)}{q} =  \frac{\mathrm{tr}(A_{-i}B_{-i})}{q^{N-1}} \mdelta^{ab}
\end{equation}
The terms in (\ref{eq:TinfOTOC}) for each $i$ (summed over $c$) project onto operators which are identical and act non-trivially on $i$.   At $T=\infty$, a $q$-body operator $\mathcal{O}_{i_1}\otimes \cdots \otimes \mathcal{O}_{i_q}$ is an eigenvector of $\mathcal{S}$ with eigenvalue $q$.   (\ref{eq:TinfOTOC}) is slightly modified in the presence of fermionic DOF, which anticommute on distinct sites:  see Appendix \ref{app:free}.   Similar definitions of size, applicable at $T=\infty$, are found in \cite{nahum, tibor, stanford1802}.

We are now ready to reinterpret conjecture (\ref{eq:main}).   Let $R\propto N^0$ be a user defined cutoff which separates small operators (eigenvectors of $\mathcal{S}$ with eigenvalue $\le R$) from large operators (eigenvalue $>R$).   Let $\mathfrak{p}$ denote a projector onto small operators, and let $\mathfrak{p}|A) = |A)$ be a typical small operator.\footnote{A typical operator refers to one which has no overlap with conserved quantities such as $H$, or long wavelength hydrodynamic modes.}   Defining $\tau$ as the time required for operator $|A)$ to grow large: \begin{equation}
(A(s) | \mathfrak{p}|A(s)) < \delta (A|A) \;\;\; (s\ge \tau),  \label{eq:taudef}
\end{equation}
where $0<\delta < 1$ is another user defined constant,  we conjecture that there exists an $R\propto N^0$ and $\delta \propto N^0$ for which $\tau$ obeys (\ref{eq:main}) in $k$-local quantum systems.  

While we cannot diagonalize $\mathcal{S}$ at finite temperature, we can still constrain the average size of a $q$-local operator $A$:   $\frac{(A|\mathcal{S}|A)}{(A|A)}$ is a sum of $k(q^2-1)$ thermal correlators normalized by $\chi_{AA}$.  If $q\propto N^0$, then in any extensive system obeying $\chi_{AA} \propto N^0$, $A$ has average size $\frac{(A|\mathcal{S}|A)}{(A|A)} \propto N^0$.   Furthermore, any operator with average size $<R$ must contain some small components: since $(A(t)|\mathcal{S}|A(t)) \ge (A(t)|\mathfrak{q}\mathcal{S}\mathfrak{q}|A(t)) \ge R(A(t)|\mathfrak{q}|A(t))$, \begin{equation}
\frac{(A(t)|\mathfrak{p}|A(t))}{(A|A)}  \ge 1- \frac{(A(t)|\mathcal{S}|A(t))}{R(A|A)}.  \label{eq:markov}
\end{equation}
Even at finite temperature, (\ref{eq:TinfOTOC}) ensures that ``small" $q$-local operators remain relatively small, for sufficiently large $R\propto N^0$.

\section{Evidence}
Remarkably, our conjecture is consistent with all quantum many-body systems studied to date.   Despite the abstract formulation of (\ref{eq:taudef}), there are a number of simple ways to bound $\tau$.

\begin{figure}
\centering
\includegraphics[width=5in]{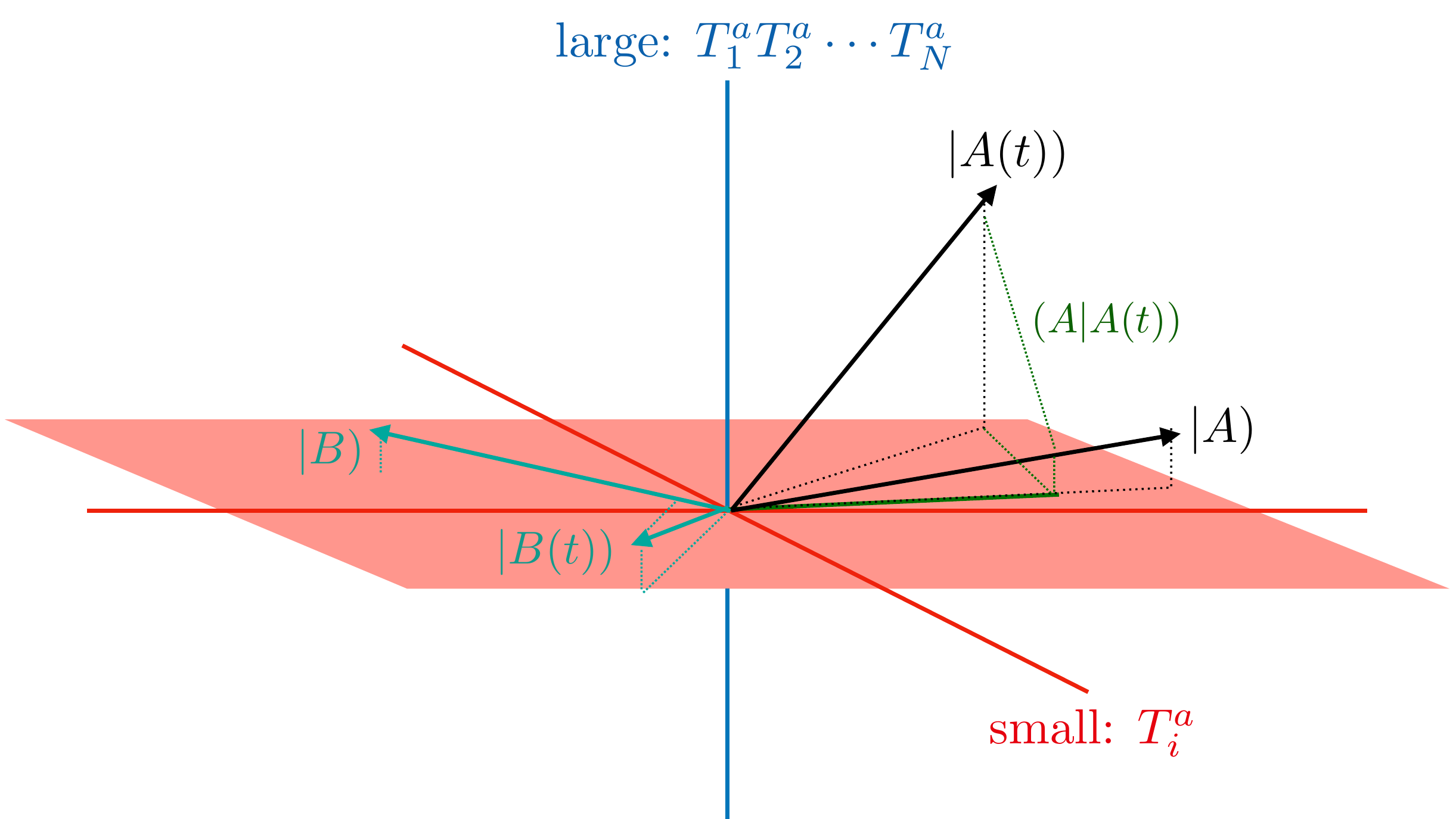}
\caption{A cartoon of operator space.  The subspace of small operators is shaded.   Since operator $|A)$ is mostly small (depicted by dotted lines), we know that $|A(t))$ has a significant small component whenever $(A|A(t))$ is sufficiently large.   There is no bound on how quickly the nearly small operator $|B)$ can evolve into an orthogonal operator in the small subspace.  Our conjecture is that (\ref{eq:main}) only constrains the time for $|B)$ to rotate into the large subspace.}
\label{fig}
\end{figure}

One bound on $\tau$ comes from calculating thermal two point functions $\langle A(t)A\rangle$, where $A$ is a $k$-local operator.   Suppose for simplicity that, even at finite $T$,  $A$ were entirely small:  $\mathfrak{p}|A) = |A)$.   Then since $\mathfrak{p}$ projects on to \emph{all} small operators, not only $A$: $(B|\mathfrak{p}|B) \ge (B|A)^2$ for any operator $B$.    Choosing operator $B=A(t)$,  we find that $(A(t)|\mathfrak{p}|A(t)) \ge (A(t)|A)^2$.   We conclude that $(A(\tau)|A) \ge \sqrt{\delta}$, which provides a physical constraint on $\tau$ arising from two point functions.   In general, a few-body operator $|A)$ is a sum of small and large operators at finite $T$.  Applying the Cauchy-Schwarz inequality to each term in $(A(t)|A) = (A(t)|\mathfrak{p}+\mathfrak{q}|A)$, along with $(A(t)|\mathfrak{q}|A(t))\le (A|A)$:
\begin{equation}
(A(t)|A) \le \sqrt{(A(t)|\mathfrak{p}|A(t)) (A|\mathfrak{p}|A)} + \sqrt{(A|\mathfrak{q}|A)(A|A)}.  \label{eq:cauchy}
\end{equation}
If $(A|\mathfrak{q}|A) \ll 1$ (which can be checked using (\ref{eq:markov})), $\tau$ remains bounded by the two point function $(A(t)|A)$.    A pictorial demonstration of this result is provided in Figure \ref{fig}.

The conjectured Planckian bound (\emph{i}) states that $(A(t)|A) \sim \mathrm{e}^{-t/\tau_2}$, where $\tau_2 \gtrsim \beta$.   Taking $\delta \ll 1$, we conclude that $\tau \gtrsim \tau_2 \log \frac{1}{\delta}$.   Our conejecture implies bound (\emph{i}).  As all strongly coupled systems studied to date satisfy bound (\emph{i})), they also satisfy our conjectured bound.  An explicit example of operator size dynamics in a strongly coupled system (the Sachdev-Ye-Kitaev model \cite{sachdevye, stanford1604, suh}) is  in Appendix \ref{app:syk}.   

More generally, we argue in Appendix \ref{app:random} that in extensive, spatially local theories, the decay time $\tau_2^*$ of a random operator in a thermodynamically large region of entropy $S_{\mathrm{th}}$ obeys $\tau_2^* \gtrsim \beta/\sqrt{S_{\mathrm{th}}}$.  This is quite similar to (\ref{eq:main}), albeit with a large value of $R$ in (\ref{eq:taudef}), and a possibly small prefactor in the Planckian bound.   We derive this bound by relating the annealed average of $\frac{(A(t)|A)}{(A|A)}$ over random operators $A$ to the thermal partition function $Z(\beta) = \mathrm{tr}(\mathrm{e}^{-\beta H})$, evaluated at complex temperature.  

Another bound on $\tau$ comes from quantum many-body chaos.  In a non-integrable system, we expect that for a random pair of few-body operators $A$ and $B$,  the out-of-time-ordered correlator (OTOC) $\langle A(t)BA(t)B\rangle$ starts off as $\approx \langle AA\rangle\langle BB\rangle$ at $t\lesssim \beta$, and shrinks to 0 as $t\rightarrow \infty$ \cite{bhbutterfly}.   For our purposes, it is more convenient to re-cast the OTOC as the growth of the squared commutator $-\langle [A(t),B]^2\rangle$.   Sending $B\rightarrow T^a_i$ and summing over $a,i$ as in (\ref{eq:TinfOTOC}), the sum of OTOCs is simply $(A(t)|\mathcal{S}|A(t))$.   Chaos corresponds to the growth in OTOCs, and thus in average operator size.   Heuristically, the formal bound on chaos (\emph{v}) states that $\langle [A(t),B]^2\rangle \lesssim \frac{1}{N} \mathrm{e}^{2\mpi T t}$ whenever $\langle [A(t),B]^2\rangle \lesssim \frac{1}{N}$ for $t\lesssim \beta$ \cite{stanfordbound}.    Within our framework, this bound admits a more physical interpretation: if $A$ is (mostly) small, as is $A(t)$ for all $t\lesssim \beta$,  then the Planckian time bounds the growth of $(A(t)|\mathcal{S}|A(t))$ at later times.  Hence, our conjecture is one of the postulates required for the chaos bound (\emph{v}).  A more technical discussion is presented in Appendix \ref{app:chaos}.

\section{Coupling Strength}
Our conjectured bound (\ref{eq:main}) is never saturated in integrable models, whose Hamiltonian $H_{\mathrm{int}}$ has arbitrarily many (quasi)local conserved charges $\lbrace Q_a\rbrace $ \cite{prosen15}, for which $(Q_a|\mathcal{S}|Q_a) \propto N^0$ in the thermodynamic limit.\footnote{More precisely, for any fixed number $R$, we expect at least $\mathrm{O}(R^\alpha)$ conserved quantities $Q_a$ for which $(Q_a|\mathcal{S}|Q_a) \le R$.   In the Heisenberg model in one dimension, $\alpha=1$ \cite{prosen15}.}   After all, $Q_a(t)=Q_a$ remains small for all $t$; thus $\tau=\infty$.  An explicit example is a theory of free fermions perturbed by interactions: \begin{equation}
H = \sum_i \epsilon_i c^\dagger_i c_i +  \lambda \sum_{ijkl} U_{ijkl}c^\dagger_i c^\dagger_j c_k c_l   \label{eq:freefermions}
\end{equation}
with $\lambda$ perturbatively small.   Here $\lbrace c_i^\dagger, c_j\rbrace = \mdelta_{ij}$ are conventional creation/annihilation operators.  When $\lambda=0$, the operator $c_i^\dagger c_i$ does not evolve with time.   Using Fermi's golden rule, we estimate that $\langle (c_i^\dagger c_i)(t) c_i^\dagger c_i \rangle \gtrsim \mathrm{e}^{-t/\tau_2} $, where the decay rate $\tau_2^{-1} \propto \lambda^{2}$.    From (\ref{eq:cauchy}), we conclude (\ref{eq:main}) holds for small enough $\lambda$. 
Moreover, if the coupling $\lambda$ is irrelevant, then as $T\rightarrow 0$, $\tau_2$ diverges as $T\rightarrow 0$ even at finite $\lambda$.

While operators cannot grow large quickly as $\lambda \rightarrow 0$, there is no universal bound (\emph{i}) on the decay of two point functions.  Consider the free fermion Hamiltonian $H = h_{ij} c^\dagger_i c_j$, where $h_{ij}$ consists of hopping on a three dimensional lattice with weak on-site disorder.  Then $\langle c^\dagger_{\mathbf{k}}(t) c_{\mathbf{k}}\rangle \sim \mathrm{e}^{-t/\tau_{\mathrm{imp}}}$ for a plane wave state $c_{\mathbf{k}} = \sum \mathrm{e}^{\mathrm{i}\mathbf{k}\cdot \mathbf{x}_i} c_i$ \cite{palee}.   $\tau_{\mathrm{imp}}$ is a disorder scattering time obtained by finding the quantum motion of one particle, and is manifestly independent of $T$.  There is no Planckian bound on $\tau_{\mathrm{imp}}$ because one small operator $c_i(t) = G_{ij}(t)c_j$ can decay exponentially  into other small operators $c_j$, and $G_{ij}(t)$ solves a state independent equation:  $\frac{\mathrm{d}}{\mathrm{d}t} G = \mathrm{i}[h,G]$.   Any operator $c_j$ at the Fermi surface will be detected in $(c_i(t)|c_i)$; even as $T\rightarrow 0$ exponential decay is allowed.  

Choosing $h_{ij}$ to instead be a lattice discretization of Dirac fermions with speed of light $c$,  near the light cone $ct = |\mathbf{x}_i-\mathbf{x}_j|$ the OTOC $\langle c^\dagger_i(t) c_j c^\dagger_i(t) c_j \rangle$ is sharply varying and does not obey the chaos bound \cite{stanfordbound}.   However, we show in Appendix \ref{app:free} that $\frac{\mathrm{d}}{\mathrm{d}t} (A(t)|\mathcal{S}|A)=0$ for arbitrary operators $A$ in a free theory.    Individual terms in the sum (\ref{eq:TinfOTOC}) which defines $(A(t)|\mathcal{S}|A(t))$ are unbounded;  only their sum is constrained.

If (\ref{eq:main}) represents a bound on thermalization, $\tau$ should diverge in any integrable theory.  Hence, neither local two-point functions nor OTOCs always obey Planckian bounds.  In contrast, if $\tau$ is the time over which small operators become large, $\tau=\infty$ in integrable systems, and $\tau$ remains large at perturbatively large at weak coupling.   Integrability must be broken by a sufficiently large amount in order to saturate (\ref{eq:main}).  Consistent with physical intuition that thermalization should be bounded only at strong coupling, our conjectured bound (\ref{eq:main}) is only saturated sufficiently far from $\lambda = 0$.

In special models, such as the quantum Ising model on a one dimensional lattice \cite{sachdev}, $H$ looks $k$-local under two ``dual" decompositions of the Hilbert space, related by a nonlocal transformation.   In these models, in a non-interacting basis where $H=h_{ij}c^\dagger_i c_j$, the average size of small operators such as $c_i(t)$ is time independent;  in an interacting basis, $H$ is local in terms of operators such as $V_n = \exp[\mathrm{i} \sum_{j<n}  c_j](t)$, which can algebraically  grow large \cite{dora, motrunich}.    In both bases, our conjecture (\ref{eq:main}) is obeyed.

\section{Transport Bounds}
 
Next, we discuss transport bounds (\emph{ii})-(\emph{iv}),  focusing for simplicity on bounds on electrical conductivity $\sigma = \chi_{JJ} \tau_{\mathrm{tr}}$ in isotropic systems (other transport bounds have similar subtleties).    Here $\chi_{JJ} = \beta (J_x|J_x)$ is the current-current susceptibility, and \cite{lucasbook} \begin{equation}
\tau_{\mathrm{tr}} = \int\limits_0^\infty \mathrm{d}t \; \frac{(J_x|J_x(t))}{(J_x|J_x)}.  \label{eq:tautrdef}
\end{equation}

A Planckian bound on $\tau_{\mathrm{tr}}$ is not general:   with weak short-range disorder, non-interacting fermions have a residual $\tau_{\mathrm{tr}} \approx \tau_{\mathrm{imp}}$ \cite{palee} at $T=0$.    In fact, $\tau_{\mathrm{tr}}$ is generally unphysical: it cannot be measured by the location of a pole or branch cut in any correlation function.      A simple example is transport in a weakly disordered metal with quadratic dispersion relation in two spatial dimensions.   In a weak background magnetic field, defining $j_i  =(J_i|J_x(t))$:   \cite{lucasMM}
    \begin{equation}
\frac{\mathrm{d}}{\mathrm{d}t} \left(\begin{array}{c}  j_x \\ j_y  \end{array}\right)  = -\left(\begin{array}{cc}  \tau_{\mathrm{imp}}^{-1} &\ \omega_{\mathrm{c}} \\  -\omega_{\mathrm{c}} &\ \tau_{\mathrm{imp}}^{-1} \end{array}\right)  \left(\begin{array}{c}  j_x \\ j_y  \end{array}\right) \equiv - \Gamma \left(\begin{array}{c}  j_x \\ j_y  \end{array}\right) .  \label{eq:jxjy}
\end{equation}
where $\omega_{\mathrm{c}}$ is the cyclotron frequency.  Combining (\ref{eq:tautrdef}) and (\ref{eq:jxjy}):  
\begin{equation}
\tau_{\mathrm{tr}} = \frac{(J_x| \Gamma^{-1} |J_x)}{(J_x|J_x)^2},  \label{eq:Gamma1}
\end{equation}
and in this example $\tau_{\mathrm{tr}} = \tau_{\mathrm{imp}}/(1+(\omega_{\mathrm{c}}\tau_{\mathrm{imp}})^2)$.    Since $\langle J_x(t)J_x\rangle \sim \cos(\omega_{\mathrm{c}}t) \mathrm{e}^{-t/\tau_{\mathrm{imp}}}$,  $\tau_{\mathrm{tr}}$ does not control the decay of the current operator.   Even if $\tau_{\mathrm{imp}}$ is bounded, $\tau_{\mathrm{tr}}$ can take any non-negative value by tuning $\omega_{\mathrm{c}}$ via magnetic field (an external, unbounded coupling constant).    (\ref{eq:Gamma1}) is not inconsistent with (\ref{eq:main}) because $\tau_{\mathrm{tr}}$ is sensitive to both dissipative time scales such as $\tau_{\mathrm{imp}}$, and  non-dissipative time scales such as $\omega_{\mathrm{c}}^{-1}$.

Using the memory matrix formalism \cite{lucasbook}, we formally show in Appendix \ref{app:MBL} that (\ref{eq:Gamma1}) admits a natural generalization, where $\Gamma$ is a matrix whose indices correspond to each small operator.   The antisymmetric part of $\Gamma$ describes rotation of small operators among themselves, analogous to $\omega_{\mathrm{c}}$;  the symmetric part describes the decay of small operators into large operators ($\tau^{-1}_{\mathrm{imp}}$).    Only the symmetric part of $\Gamma$ is constrained in any way by (\ref{eq:main}).   There is no bound on $\tau_{\mathrm{tr}}$, in general.

Interplay between small operator rotation vs. growth/dissipation in (\ref{eq:Gamma1}) is the physical mechanism which allows disorder-driven metal-insulator transitions (MIT) to avoid prior (Planckian) transport bounds,  both near a non-interacting Anderson transition and near many-body localized (MBL) phases.   In these localized phases, we show in Appendix \ref{app:MBL} that $|J_x)$ overlaps only with the antisymmetric part of $\Gamma^{-1}$:  hence from (\ref{eq:Gamma1}),  $\tau_{\mathrm{tr}}=0$.    The conductivity vanishes because small operators simply rotate among themselves:  there is no dissipation and no charge transport.   While $\tau_{\mathrm{tr}}$ does not obey (\ref{eq:main}), the decay time for small operators $\tau$ does.  This can be seen in a fully many-body localized (MBL) phase \cite{serbyn}.\footnote{This phase appears to  be the most disordered of a sequence of phase transitions driven by disorder \cite{demler2, demler, varma, swingle1807}; the MIT is the first of such transitions.}  A model Hamiltonian, with local Hilbert space dimension $q=2$, is $H = h_i \sigma_i^z + h_{ij} \sigma_i^z \sigma_j^z + h_{ijk} \sigma_i^z \sigma_j^z \sigma_k^z +  \cdots$, with $h_{ij\cdots}$ random couplings which exponentially decay with the distance between $ij\cdots$ \cite{serbyn}.   $|J_x)$ is a local operator.  Since $H$ contains only $\sigma^z$, operator dynamics is extremely slow:  only exponentially suppressed couplings between distant $i$ and $j$ can grow $|J_x(t))$.   Hence  $R \propto \log \tau$, and (\ref{eq:main}) is obeyed \cite{xie2017, fan2017out, chenlogarithm, debanjan2}.

We expect $\tau_{\mathrm{tr}}$ is also unphysical in strongly coupled MITs, including in holographic\footnote{Strictly speaking, it may be the case that holographic models of the MIT are not $k$-local (and so our results would not apply).  However, the simplest quantum theories with some kind of holographic dual description are $k$-local, including the SYK model \cite{sachdevye, stanford1604, suh} and matrix models \cite{bfss}.}  models \cite{donos12, gouteraux2, mefford, donos14} where $\sigma \rightarrow 0$:  correlators whose decay times obey (\ref{eq:main}) exist in all known holographic models, implying that small operators do not grow large at the MIT.   The insulating transition is possible because $\tau_{\mathrm{tr}}$ need not relate to (\ref{eq:main}) if small operators can rotate amongst themselves.   Interplay between dissipative and non-dissipative time scales in (\ref{eq:Gamma1}) plausibly resolves all loopholes to transport bounds (\emph{ii})-(\emph{iv}).

\section{Outlook}
We conjecture that in generic $k$-local systems, (\ref{eq:main}) is obeyed if $\tau$ is the decay time of small operators.  This bound improves earlier conjectures and is consistent with all known many-body quantum systems.  Our conjecture is testable experimentally by measurements of thermal correlators, including those with unusual time ordering \cite{garttner, du, cappellaro}, or by other probes of operator dynamics \cite{ana}.

Verifying this conjecture will fundamentally constrain thermalization in quantum systems and resolve many open questions.   In particular, we expect that the defining characteristic of a strongly coupled quantum system is the rapid and direct decay of small operators into large operators, as measured in any basis where the Hamiltonian is local.    It is in this limit that measurable time scales, such as the decay times of two point functions, or transport times, may be comparable to the decay time of small operators.   This explains why the ``saturation" of Planckian transport and chaos bounds is a generic feature of strongly coupled systems, even when such bounds do not apply to weakly coupled systems.  

\addcontentsline{toc}{section}{Acknowledgements}
\section*{Acknowledgements}
I thank Yingfei Gu, Sean Hartnoll, Xiao-Liang Qi, Koenraad Schalm and Alex Streicher for helpful discussions.  I thank NORDITA for hospitality during the program ``Bounding Transport and Chaos", along with the participants for useful feedback.  I am supported by the Gordon and Betty Moore Foundation's EPiQS Initiative through Grant GBMF4302.

\begin{appendix}

\section{Free Fermions}\label{app:free}
In this appendix, we consider operator size and dynamics in a theory of (spinless) free fermions with Hamiltonian \begin{equation}
H = h_{ij}c^\dagger_i c_j  \label{eq:HappB}
\end{equation}
where $h_{ij}$ is an arbitrary Hermitian matrix.   We employ Einstein summation conventions on indices.

\subsection{Eigenstates and the Current Susceptibility}
Let us first define the unitary transformation $U$:  \begin{equation}
c_j  = U_{j\alpha}c_\alpha  \label{eq:unitary}
\end{equation}
to the eigenbasis $\alpha$ of single particle eigenstates where (\ref{eq:HappB})  becomes \begin{equation}
H = \sum_\alpha \epsilon_\alpha c^\dagger_\alpha c_\alpha.
\end{equation}
The conductivity is a fermion bilinear operator of the form $J_{ij}c^\dagger_i c_j$ where $J$ is a Hermitian matrix.   A natural basis for such operators is \begin{subequations}\begin{align}
|\alpha\alpha) &=  c^\dagger_\alpha c_\alpha - f_\alpha, \\
 |\alpha\beta)_- &=  \frac{c^\dagger_\alpha c_\beta - c^\dagger_\beta c_\alpha }{\sqrt{2}\mathrm{i}}, \\
  |\alpha\beta)_+ &=  \frac{c^\dagger_\alpha c_\beta + c^\dagger_\beta c_\alpha}{\sqrt{2}},
\end{align}\end{subequations}
where \begin{equation}
f_\alpha = \frac{\mathrm{e}^{-\beta\epsilon_\alpha}}{1+\mathrm{e}^{-\beta\epsilon_\alpha}}.
\end{equation}
The inner product (\ref{eq:innerproduct}) on such operators is \begin{subequations}\begin{align}
(\alpha\alpha|\beta\beta) &= \mdelta_{\alpha\beta}\times f_\alpha(1-f_\alpha), \\
\;_{\sigma}(\alpha\beta|\gamma\mdelta)_{\sigma^\prime} &= \mdelta_{\sigma\sigma^\prime} (\mdelta_{\alpha\gamma}\mdelta_{\beta\delta} + \mdelta_{\alpha\delta} \mdelta_{\beta\gamma}) \times  \frac{f_\alpha-f_\beta}{\beta(\epsilon_\beta - \epsilon_\alpha)}
\end{align}\end{subequations}
where $\sigma$ denotes $\pm$.   

If the current operator is $J = J_{\alpha\beta}c^\dagger_\alpha c_\beta$, then \begin{equation}
\chi_{JJ} = \frac{(J|J)}{T} = \sum_{\alpha,\beta} |J_{\alpha\beta}|^2 \frac{f_\alpha-f_\beta}{\epsilon_\beta - \epsilon_\alpha}
\end{equation}
In a free theory $J_{\alpha\alpha} \ne 0$, though with disorder we expect that $J_{\alpha\alpha}=0$ (as particles should either diffuse or be localized).   So long as there exists a sufficient density of states $\alpha$ and $\beta$ with $\epsilon_\alpha < 0 < \epsilon_\beta$ and $J_{\alpha\beta}\ne 0$, $\chi_{JJ}>0$ approaches a $T$-independent positive constant as $T\rightarrow 0$.  Even in the Anderson localized phase, we expect such pairs of localized states will generically exist.

More generally, a natural basis for all possible operators consists of a tensor product of operators $1$, $c_\alpha$, $c_\alpha^\dagger$, $c^\dagger_\alpha c_\alpha - f_\alpha$.    Because the density matrix of free fermions can be written as $\rho = \bigotimes \rho_\alpha$,   it is easy to show that the four operators listed above are orthogonal, and thus generate a complete orthogonal basis of operators.  Note that unlike the real vector space of Hermitian operators described in the main text, this is a complex vector space which includes non-Hermitian operators.

\subsection{Operator Size}
Now let us study the dynamics of operator size in a  theory of free fermions.   We will use a slightly modified definition of $\mathcal{S}$ as compared to (\ref{eq:TinfOTOC}) which is more naturally suited to a theory of fermions:   \begin{equation}
(A|\mathcal{S}|A) =  \sum_i  \left[ ([c_i, A]_\eta |[c_i,A]_\eta )  + ([c_i^\dagger, A]_\eta |[c_i^\dagger,A]_\eta ) \right]\label{eq:sizefermion}
\end{equation}
where $\eta=1$ denotes the commutator if the operator $A$ contains an even number of fermions (is bosonic) and $\eta=-1$ denotes the anticommutator if $A$ contains an odd number of  fermions.    The reason for this choice is that a product of a small number of fermion operators, such as $c_1 c_2 c_3$, should have an $N$-independent size.   We can ensure this is true (in an extensive quantum system, at any temperature) by replacing the commutator with the graded commutator when defining $\mathcal{S}$.   %Since $A$ is Hermitian, we can use either $c_i$ or $c_i^\dagger$ in (\ref{eq:sizefermion}).

%If we had instead used (\ref{eq:TinfOTOC}), the sum over commutators with $c_i$ in (\ref{eq:sizefermion}) would be replaced with a sum over (suitably normalized)  commutators with the three operators $\mathrm{i}(c_i-c_i^\dagger)$, $(c_i+c_i^\dagger)$, and $c^\dagger_ic_i$.    In general, (\ref{eq:HappB}) no longer consists of a sum of operators of the same  size, which is why we expect (\ref{eq:sizefermion}) to be a better definition of $\mathcal{S}$ in a fermionic theory. 

One very useful property of (\ref{eq:sizefermion}) is that it is ``basis independent".   Consider the Hermitian matrix \begin{equation}
\mathcal{S}_{ij}(A) = ([c_i,A]_\eta |[c_j,A]_\eta ) + ([c_i^\dagger,A]_\eta |[c_j^\dagger,A]_\eta ).
\end{equation}
Using (\ref{eq:sizefermion}), we conclude that the size of the operator $A$ is given by $\mathrm{tr}(\mathcal{S}(A))$.   Using (\ref{eq:unitary}), we see \begin{align}
\mathcal{S}_{\alpha\beta} &= ([c_\alpha,A]_\eta |[c_\beta,A]_\eta ) + ([c^\dagger_\alpha,A]_\eta |[c^\dagger_\beta,A]_\eta ) \notag \\
&= (U^\dagger)_{\alpha i} ([c_i,A]_\eta |[c_j,A]_\eta ) U_{j\beta} + U_{\alpha i} ([c_i^\dagger,A]_\eta |[c_j^\dagger,A]_\eta ) (U^\dagger)_{j\beta},
\end{align}
i.e. the matrix $\mathcal{S}$ consists of a sum of two terms, each of which transforms in a straightforward way.   Upon evaluating the trace $\mathcal{S}_{\alpha\alpha}$, we see that the factors of $UU^\dagger$ cancel;  thus operator size is the same in each single-particle basis.  It is natural to work in the eigenbasis of $H$.

Now consider a generic operator in the eigenstate basis:  \begin{equation}
\mathcal{O} = \sum_{\sigma_\alpha, \sigma_\alpha^\prime} C_{\sigma_\alpha \sigma_\alpha^\prime} \mathcal{O}_{\sigma_\alpha\sigma_\alpha^\prime}
\end{equation} 
where \begin{equation}
\mathcal{O}_{\sigma_\alpha\sigma_\alpha^\prime} = \prod_{\alpha} \left(c^\dagger_\alpha \right)^{\sigma_\alpha^\prime} \left(c_\alpha\right)^{\sigma_\alpha}.
\end{equation}
Since \begin{equation}
\mathcal{O}(t) =  \sum_{\sigma_\alpha, \sigma_\alpha^\prime} C_{\sigma_\alpha \sigma_\alpha^\prime} \exp\left[ \mathrm{i}t\sum_\alpha \epsilon_\alpha (\sigma_\alpha-\sigma_\alpha^\prime)  \right] \mathcal{O}_{\sigma_\alpha\sigma_\alpha^\prime}
\end{equation} 
we conclude that
\begin{equation}
(\mathcal{O}(t)|\mathcal{S}|\mathcal{O}(t)) = \sum_\beta \sum_{\sigma_\alpha, \sigma_\alpha^\prime} |C_{\sigma_\alpha \sigma_\alpha^\prime}|^2 \left[ ([c_\beta, \mathcal{O}_{\sigma_\alpha\sigma_\alpha^\prime}]_\eta | [c_\beta, \mathcal{O}_{\sigma_\alpha\sigma_\alpha^\prime}]_\eta) + ([c_\beta^\dagger, \mathcal{O}_{\sigma_\alpha\sigma_\alpha^\prime}]_\eta | [c_\beta^\dagger, \mathcal{O}_{\sigma_\alpha\sigma_\alpha^\prime}]_\eta) \right] \label{eq:constantfreesize}
\end{equation}
Note that there are no further cross terms relative to what was written above due to the orthogonality of operators when written in the eigenbasis.   Since average operator size does not grow in a fermionic theory, we conclude that the decay rate for small operators is infinite, as claimed in the main text.

 \section{The SYK Model} \label{app:syk}
We now describe operator dynamics in the Sachdev-Ye-Kitaev (SYK) model, which is a $q$-local ($q>2$) model of $N$ Majorana fermions $\chi_i$ ($i=1,\ldots,N$)   \cite{sachdevye, stanford1604, suh} obeying $\lbrace \chi_i, \chi_j \rbrace = \mdelta_{ij}$:  
 \begin{equation}
 H = \sum_{i_1<i_2<\ldots <i_q}  J_{i_1\cdots i_q} \chi_{i_1}\cdots \chi_{i_q},
 \end{equation}
 where $q$ is even, $J_{i_1\cdots i_q} $ are Gaussian random coupling constants with variance \begin{equation}
 \overline{J_{i_1\cdots i_q}^2} = \frac{2^{q-1}(q-1)!}{q} \frac{J^2}{N^{q-1}}. 
 \end{equation}
For simplicity in this appendix, we take \begin{equation}
(A|B) = \mathrm{tr}(\sqrt{\rho}A^\dagger\sqrt{\rho}B) \label{eq:SYKsize2}
\end{equation}
 and define the size operator as
\begin{equation}
(A|\mathcal{S}|B) = \sum_{i=1}^N ([ A,\chi_i]_\eta  | [ B,\chi_i]_\eta).  \label{eq:SYKsize}
\end{equation}
Here $[,]_\eta$ denotes the graded commutator, as before.   

  At $T=\infty$ and at large $q$,  operator dynamics was studied in \cite{stanford1802}:  one finds that $(\chi_1(t)|\chi_1) \approx \mathrm{e}^{-2Jt/q}$ while 
$(\chi_1(t)|\mathcal{S}|\chi_1(t)) \approx \frac{1}{2} \mathrm{e}^{2Jt} + \cdots $; here and below $\cdots$ denotes subleading contributions at large $t$.    Up to the factor of $\frac{1}{2}$, these two formulas are identical to what is predicted by classical infection dynamics \cite{nahum, tibor, stanford1802}:  operators grow large because their small constituents decay into $\approx q$ larger operators.  
  
  When $\frac{1}{N} \ll T\ll J$,  (\ref{eq:mss}) becomes saturated while $(\chi_1(t)|\chi_1) \sim (\beta J)^{-2/q} \mathrm{e}^{-2\mpi t/\beta q}$ \cite{stanford1604}, which is suggestive  of the classical infection analogy also holding at finite $T$.  We find that $(\chi_1(t)|\mathcal{S}|\chi_1(t)) \sim (\beta J)^{-4/q} \mathrm{e}^{t/\tau_{\mathrm{L}}} + \cdots$ with $\tau_{\mathrm{L}} = \frac{\beta}{2\mpi}(1+\frac{2}{\beta J}+\cdots )$.  Since $(\chi_1| \chi_1) \sim (\beta J)^{-2/q}$ \cite{stanford1604}, we conclude that a single fermion operator has size $\sim (\beta J)^{2/q}$.   As $q\rightarrow \infty$, $(\beta J)^{2/q}\rightarrow 1$ and the $T$-dependent enhancement of size is unimportant.
  
Calculating more generic OTOCs, one finds that \cite{kitaev2017}
\begin{equation}
\sum_j \mathrm{tr}(\sqrt{\rho} \lbrace \chi_i(t), \chi_j\rbrace  \sqrt{\rho} [\chi_i(t),\chi_j]) \sim \beta J (\chi_1(t)|\mathcal{S}|\chi_1(t))   + \cdots.
\end{equation}   The specific operator ordering of (\ref{eq:SYKsize}) leads to a cancellation of the $\beta J$ enhancement in a more generic OTOC.   Interestingly, this cancellation leads to an infection-like analogy for operator dynamics at finite $T$ at large $q$, with operator growth occurring on the time scale $\tau_{\mathrm{L}}\sim \beta$ and operator decay simultaneously occuring over a time $\tau_2\sim q\beta$.   At finite $q$, there is an apparent ``lag" in the exponential growth of chaos, as compared to the infection analogy, implying that $\chi_1(t)$ is also decaying into other small operators.   At any $q$, the SYK model is consistent with the notion that small operators do not decay faster than allowed by the Planckian rate.    

The SYK  model shares many common features with quantum gravity in nearly-$\mathrm{AdS}_2$ spacetimes \cite{stanford1604, suh, jensen, maldacena16}.   There are alternative notions for measuring the effective size of operators at finite $T$ in the SYK model \cite{streicher} which may have an elegant dual interpretation in the gravity theory  \cite{lenny1, lenny2}.  A bound (\ref{eq:main}) on the lifetime of small operators, using the definition of size from \cite{streicher}, requires a much larger cutoff $R$ than using our definition.

\section{Decay of a Random Operator}\label{app:random}
Here we calculate the decay of a random operator in a many-body quantum system.   Similar calculations to the one above, in the context of random matrix theory, are found in \cite{yoshida17}.  

More precisely, we will calculate $(A(t)|A)$ using the same finite temperature inner product (\ref{eq:SYKsize2}) that we used for the SYK model in Appendix \ref{app:syk}.   The average over operators is taken to be uniform over all Hermitian operators acting on the many-body Hilbert space of fixed infinite temperature norm.   A useful basis for all such operators is given in (\ref{eq:eigenstates}).   We find that \begin{equation}
\mathbb{E}_{\mathrm{op}}\left[ (A(t)|A)\right] = \frac{1}{q^{2N}Z(\beta)}\sum_{\alpha\beta} \mathrm{tr} \left[|\alpha\rangle\langle \beta| \mathrm{e}^{-(\frac{\beta}{2}+\mathrm{i}t)H}|\beta\rangle\langle \alpha| \mathrm{e}^{-(\frac{\beta}{2}-\mathrm{i}t)H}\right] = \frac{1}{q^{2N}Z(\beta)} \left|Z\left(\frac{\beta}{2}-\mathrm{i}t\right)\right|^2.  \label{eq:Eop}
\end{equation}
where $\mathbb{E}_{\mathrm{op}}[\cdots]$ denotes the uniform average over operators described above and $Z$ is the partition function analytically continued to complex temperature.  It is also straightforward to remove all operators $|\alpha\alpha)$ from the average (which we denote as $\mathbb{E}^\prime_{\mathrm{op}}[\cdots]$).   This may be useful since $|\alpha\alpha)$ are non-dynamical and trivially commute with the Hamiltonian $H$.\footnote{Of course, in the thermodynamic limit, there are many further approximate degeneracies that will appear.  These will imply the existence of many-body operators with extremely slow dynamics.  We will show that this partial subtraction is still useful as a reference point for understanding the time scale at which an average non-trivial operator can decay.}  This leads to a slightly improved formula:  \begin{align}
\mathbb{E}^\prime_{\mathrm{op}}\left[ (A(t)|A)\right] &= \frac{q^N}{q^N-1} \left(\mathbb{E}_{\mathrm{op}}\left[ (A(t)|A)\right]  - \frac{1}{q^{2N}}\sum_{\alpha} \mathrm{tr} \left[|\alpha\rangle\langle \alpha| \mathrm{e}^{-(\frac{\beta}{2}+\mathrm{i}t)H}|\alpha\rangle\langle \alpha| \mathrm{e}^{-(\frac{\beta}{2}-\mathrm{i}t)H}\right]\right) \notag \\
& = \frac{1}{q^N(q^N-1) Z(\beta)}\left( \left|Z\left(\frac{\beta}{2}+\mathrm{i}t\right)\right|^2 - Z(\beta)\right).
\end{align}
Finally, in order to compare with our definition of the time $\tau$ which we conjecture obeys (\ref{eq:main}), it is useful to calculate $\mathbb{E}^\prime_{\mathrm{op}}[\frac{(A(t)|A)}{(A|A)}]$.  Unfortunately, this is rather difficult to average over, but we can compute a simpler ``annealed average" \begin{equation}
f(t) \equiv \frac{\mathbb{E}^\prime_{\mathrm{op}}\left[ (A(t)|A)\right] }{\mathbb{E}^\prime_{\mathrm{op}}\left[ (A|A)\right] } = \frac{\left|Z\left(\frac{\beta}{2}+\mathrm{i}t\right)\right|^2 - Z(\beta)}{Z\left(\frac{\beta}{2}\right)^2-Z(\beta)}.   \label{eq:ftdef}
\end{equation} 

It is useful to now consider an explicit example for $Z(\beta)$.  We consider quantum systems whose  low energy effective theory is accurately modeled by a scaling theory, and in which the free energy \begin{equation}
F(\beta) = -\frac{\log Z(\beta)}{\beta} = -C \beta^{-\gamma}.
\end{equation}  
Depending on the precise critical exponents \cite{lucasbook}, the value of $\gamma$ changes.  Using the thermodynamic relation $S = -\partial_T F$ and $T=\beta^{-1}$, the entropy $S$ of the quantum theory is given by
\begin{equation}
S = \gamma C \beta^{1-\gamma}.  \label{eq:Sgamma}
\end{equation}  
Assuming the third law of thermodynamics $S(\infty) = 0$, we find  $\gamma > 1$.  Typically $\gamma$ is independent of $N$.  The entropy $S$ is a crude proxy for the number of quantum degrees of freedom in our many-body system.  In our scaling theory, we therefore estimate \begin{align}
f(t) &\approx \frac{ \left|\mathrm{e}^{S(\mathrm{i}t+\beta/2)/\gamma}  \right|^2- \mathrm{e}^{S(\beta)/\gamma}}{\mathrm{e}^{2S(\beta/2)/\gamma} - \mathrm{e}^{S(\beta)/\gamma}} \approx \exp\left[ \frac{2}{\gamma} \mathrm{Re}\left(S\left(\frac{\beta}{2}+\mathrm{i}t\right)\right)-\frac{2}{\gamma} S\left(\frac{\beta}{2}\right) \right] \notag \\
&\approx \exp\left[-\frac{2}{\gamma} S\left(\frac{\beta}{2}\right) \left( \frac{\gamma(\gamma-1)}{2}  \left(\frac{t}{\beta}\right)^2 + \mathrm{O}\left(t^4\right)\right) \right].
\end{align}
Our bound on the time $\tau$ at which a typical operator could possibly grow large is therefore \begin{equation}
\tau \gtrsim \frac{\beta}{\sqrt{(\gamma-1)S}}.
\end{equation}

Clearly in a many-body system with $S\propto N$, the time scale $\beta/\sqrt{S}$ is far too small to be relevant for (\ref{eq:main}).   Nevertheless, we see that the time scale over which individual operators decay scales at least linearly with $\beta$ (if $N$ is held fixed and $S$ vanishes as $\beta \rightarrow \infty$ as in (\ref{eq:Sgamma}), we find $\tau \sim \beta^{(1+\gamma)/2}$).   The scaling of $\tau$ with $\beta$ is thus consistent with (\ref{eq:main}).

\subsection{Spatial Locality}

Next we consider spatially local theories  in $d$ dimensions.   In such models, define the region $B$ to be lattice sites contained within a ball of radius $r$, and take $r$ to be larger than the thermal correlation length.   Then it is natural to expect that at temperature $\beta$, the Hamiltonian $H\approx H_B+ H_{B^{\mathrm{c}}} + H_{\mathrm{bnd}}$, where $H_A$ is a Hamiltonian acting entirely in region $B$, $H_{B^{\mathrm{c}}} $ acts entirely in the complement of $B$, and $H_{\mathrm{bnd}}$ is a boundary operator.    Let $R$ denote the number of DOF in $B$.

We now generalize our operator average above, and restrict only to operators acting non-trivially in region $B$.   Denoting the resulting average as $\mathbb{E}_{\mathrm{op}}^B[\cdots]$, (\ref{eq:Eop}) generalizes to \begin{equation}
\mathbb{E}^B_{\mathrm{op}}[(A(t)|A)] = \frac{1}{q^{2R}Z(\beta)} \mathrm{tr}\left(\underset{B}{\mathrm{tr}} \left(\mathrm{e}^{-(\beta/2-\mathrm{i}t) H}\right) \underset{B}{\mathrm{tr}} \left(\mathrm{e}^{-(\beta/2+\mathrm{i}t) H}\right)  \right). \label{eq:partialtrace1}
\end{equation}
  The important observation is that (as operators), the spectrum of $H_B$ scales as $a^d$ while the spectrum of $H_{\mathrm{bnd}}$ scales as $a^{d-1}$.   So it is natural to expect that \begin{equation}
  \underset{B}{\mathrm{tr}} \left(\mathrm{e}^{-\beta H}\right) \approx Z_B(\beta) \exp\left[-\beta H_{B^{\mathrm{c}}} - \beta \widetilde{H}_{\mathrm{bnd}}(\beta)\right] \label{eq:partialtrace2}
  \end{equation}
where $Z_B(\beta)$ is the partition function of Hamiltonian $H_B$ and $ \widetilde{H}_{\mathrm{bnd}}(\beta)$ is a Hamiltonian restricted to sites that are within a thermal correlation length of the traced over  region $B$.   Combining (\ref{eq:partialtrace1}) and (\ref{eq:partialtrace2}) we estimate that \begin{equation}
\mathbb{E}^B_{\mathrm{op}}[(A(t)|A)] \approx \frac{|Z_B(\beta/2 + \mathrm{i}t)|^2 Z_{B^{\mathrm{c}}}(\beta)}{q^{2R}Z(\beta)} \approx \frac{|Z_B(\beta/2 + \mathrm{i}t)|^2}{q^{2R}Z_B(\beta)}.   \label{eq:ZB}
\end{equation}
The last step above assumes that the total free energy is extensive and obeys $F\approx F_B + F_{B^{\mathrm{c}}}$.   We are not guaranteed such an equality in theories without spatial locality, such as the SYK model.

We may now analyze (\ref{eq:ZB}) identically to (\ref{eq:Eop}).  Our conclusion is that a typical random operator decays on the time scale \begin{equation}
\tau \gtrsim \frac{\beta}{\sqrt{S_B}},  \label{eq:tauSB}
\end{equation}
where $S_B \propto R$ is the thermal entropy of region $B$.  We now take the limit $N\rightarrow \infty$ with $R$ fixed, and conclude that operators in the region $B$ decay on an $N$-independent Planckian scale.  In fact, in gapless theories obeying hyperscaling, we expect
\begin{equation}
S_B \propto \left(\frac{r}{\xi}\right)^d,  \label{eq:SBxi}
\end{equation}
where $\xi$ is the thermal (correlation) length scale.   Since our argument above only relied on $a\gtrsim \xi$, (\ref{eq:tauSB}) is valid in any region large enough to obey $S_B\gtrsim 1$, in which case $\tau\gtrsim \beta$.    Spatial locality and the scaling assumption (\ref{eq:SBxi}), which hold in a wide variety of theories \cite{lucasbook},  is thus sufficient to imply (\ref{eq:main}).

\subsection{Integrability}
There is a final question which we must now address.   Our conjecture is that the Planckian bound (\ref{eq:main}) is saturated in strongly coupled systems, yet (\ref{eq:Eop}) demonstrates that (in critical theories) there is a Planckian decay time whose origin is entirely \emph{thermodynamic}.   To reconcile these two results, we now track down the origin of Planckian decay in a non-interacting free fermion model.  Following  the notation of Appendix \ref{app:free}, where $\alpha$ now denote single-particle eigenstates and $\mathcal{O}_{\sigma_\alpha \sigma_\alpha^\prime} = \prod (c^\dagger_\alpha)^{\sigma_\alpha^\prime} c_\alpha^{\sigma_\alpha}$, it is straightforward to calculate
\begin{equation}
(\mathcal{O}(t)|\mathcal{O}) = \prod_{\alpha, \sigma_\alpha} \frac{\mathrm{e}^{-(\beta/2-\mathrm{i}t)\sigma_\alpha \epsilon_\alpha t}\mathrm{e}^{-(\beta/2+\mathrm{i}t)\sigma^\prime_\alpha \epsilon_\alpha t}}{1 + \mathrm{e}^{-\beta \epsilon_\alpha}}.
\end{equation}
Averaging over all operators implies averaging over all bit strings of $\sigma_\alpha, \sigma_\alpha^\prime = 0$ or 1, and assuming (without loss of generality) that $\epsilon_\alpha \ge 0$, we find 
\begin{align}
 \frac{\mathbb{E}_{\mathrm{op}}[(\mathcal{O}(t)|\mathcal{O}) ]}{\mathbb{E}_{\mathrm{op}}[(\mathcal{O}|\mathcal{O}) ]} &= \prod_\alpha \frac{\cos(\epsilon_\alpha t) + \cosh(\epsilon_\alpha \beta/2)}{1+ \cosh(\epsilon_\alpha \beta/2)} = \frac{Z(\beta/2+\mathrm{i}t)Z(\beta/2+\mathrm{i}t)}{Z(\beta/2)^2}. \label{eq:Eopfree}
\end{align}
where \begin{equation}
Z(\beta) =\exp\left[ \int \mathrm{d}\epsilon \; \nu(\epsilon)\log\left(1+\mathrm{e}^{-\beta \epsilon}\right) \right],
\end{equation}
where in the last line, we have introduced a density of states $\nu(\epsilon)$ for the single particle system.  

 Of course, our final answer reproduces (\ref{eq:Eop}), but it is useful to consider the middle formula in (\ref{eq:Eopfree}), from which it is clear that the Planckian decay of operators is simply coming from the relative dephasing of large products of $c_\alpha$ and $c_\alpha^\dagger$.   If we consider a single operator $c_\alpha(t)$, it will not exponentially decay:  $(c_\alpha(t)|c_\alpha)$ simply oscillates in time at temperature independent frequency $\epsilon_\alpha$ (which clearly does not depend on $\beta$).    The integral over a continuum density of states (in the thermodynamic limit) makes this relative dephasing appear as an actual decay of a random operator after averaging.   This effect also occurs in the integrable quantum Ising model \cite{lucas1903}.   
 
We expect that in a chaotic system whose thermodynamics imply Planckian decay of random operators, the Planckian decay time cannot be traced to the dephasing of operators of constant size.  (Recall that any product of $c_\alpha$s and $c_\alpha^\dagger$s has constant size: see (\ref{eq:constantfreesize})).   Perhaps a more careful analysis of operator size dynamics, in the formalism of this appendix, can shed light into the conditions for the saturation of the Planckian bound (\ref{eq:main}) on the decay of small operators into larger ones.

\section{Chaos Bound at Finite Temperature} \label{app:chaos}
Here we expand upon the chaos bound (\ref{eq:mss}) at finite temperature $T$.   Let $A$ and $B$ be local (and thus mostly small) operators, and let \begin{equation}
\mathcal{C}(t) = \left\langle A(t)B\left(\frac{\mathrm{i}\beta}{4}\right)A\left(t+\frac{\mathrm{i}\beta}{2}\right)B\left(\frac{3\mathrm{i}\beta}{4}\right)\right\rangle.
\end{equation} 
Formally, the chaos bound of \cite{stanfordbound} reads as follows.   Let $t_0$ be a reference time after which \begin{equation}
\left|\left\langle A(t) B B\left(\frac{\mathrm{i}\beta}{2}\right)A\left(t+\frac{\mathrm{i}\beta}{2}\right)\right\rangle \right| \le \epsilon + ( A|A)  (B|B)  \;\;\;\;\; (t> t_0)
\end{equation}
where we again use the inner product (\ref{eq:SYKsize2}).    Then \begin{equation}
\frac{\mathrm{d}}{\mathrm{d}t} \left(( A|A)  (B|B)  - \mathcal{C}\right) \le 2\mpi T \mathrm{coth} \left(2\mpi T (t-t_0)\right) \left(\epsilon + ( A|A)  (B|B)  - \mathcal{C}(t)\right)  \;\;\;\;\; (t> t_0).  \label{eq:mssformal}
\end{equation}
The bound (\ref{eq:mss}) follows whenever $\epsilon$ can be made very small (e.g., $\epsilon \lesssim \frac{1}{N}$) for the operators $A$ and $B$, at a time $t_0$ chosen where $\mathcal{C}(t) \approx ( A|A)  (B|B) $ for all $t \lesssim t_0 + \beta$.     

 For simplicity in what follows, we choose $A$ and $B$ to be local operators on distinct DOF such that $\langle A(t)B\rangle$ is small for all times $t$.   So long as thermal correlators of nearly decoupled DOF nearly factorize, at early times $t$: \begin{equation}
\left\langle A(t)B\left(\frac{\mathrm{i}\beta}{4}\right)A\left(t+\frac{\mathrm{i}\beta}{2}\right)B\left(\frac{3\mathrm{i}\beta}{4}\right)\right\rangle  \approx \left\langle A(t)A\left(t+\frac{\mathrm{i}\beta}{2}\right)B\left(\frac{\mathrm{i}\beta}{2}\right)B\left(\mathrm{i}\beta\right)\right\rangle \approx ( A|A)  (B|B)  \label{eq:2approx}
\end{equation}  
The latter approximation holds for all time $t$, in a chaotic system, while the former approximation is only true at small times.   At finite temperature, $(A(t)|\mathcal{S}|A(t))$ is given by (\ref{eq:TinfOTOC}) and is a sum over both time-ordered and out-of-time-ordered correlators, analogous to a sum over $( A|A)  (B|B) - \mathcal{C}(t)$ for different $B$ which act on each DOF.   

Let us now justify our claim in the main text that the veracity of (\ref{eq:mss}) implicitly relies on (\ref{eq:main}).   Observe that for $t - t_0 \ll \beta$,  (\ref{eq:mssformal})  becomes 
\begin{equation}
\frac{\mathrm{d}}{\mathrm{d}t} \left(( A|A)  (B|B)- \mathcal{C}\right) \le \frac{1}{t-t_0} \left(\epsilon + ( A|A)  (B|B) - \mathcal{C}(t)\right)  \;\;\;\;\; (0< t -t_0 \ll  \beta ). 
\end{equation}
If $( A|A)  (B|B) - \mathcal{C}(t)$ is not negligible for $t-t_0 \ll \beta$,  the growth in this quantity is not controlled by temperature and could be arbitrarily fast:  there is no chaos bound.    To get the chaos bound it is important for $\mathcal{C}\approx ( A|A)  (B|B)$ for $t\lesssim \beta$.   Using (\ref{eq:2approx}), we observe that $\mathcal{C}\approx ( A|A)  (B|B)$ holds when $[A(t),B]$ is sufficiently small (as $\langle A(t)B\rangle \approx 0$ for all $t$).   The smallness of $[A(t),B]$ for most pairs $A$ and $B$ implies that $(A(t)|\mathcal{S}|A(t))$ is ``small".  To be precise, in many chaotic systems one finds $[A(t),B] \lesssim \frac{1}{\sqrt{N}}$, and thus $(A(t)|\mathcal{S}|A(t))\propto N^0$, for $t\lesssim \beta$.    Using (\ref{eq:markov}), $(A(t)|\mathfrak{p}|A(t))$ cannot be too small (for sufficiently large $R$).   Hence we arrive at (\ref{eq:main}).   

As explained in \cite{stanfordbound}, and noted in the main text, this chaos bound fails for a free Dirac fermion in 1+1 dimensions, simply because on the light cone $x\sim ct$,  the operators $A= c_x^\dagger$ and $B=c_0$ abruptly fail to anticommute:  we cannot find a small $\epsilon$ and $t_0$ such that the bound effectively holds.   Another example of a system where the chaos bound need not apply is to quantum systems defined on heterogeneous networks \cite{lucas1805}.  Analogously to how infections spread super-exponentially quickly on such networks \cite{vespignani}, there exist small operators which can grow large so quickly that $\epsilon$ becomes O(1).   An explicit example of a quantum system on a heterogeneous graph where the chaos bound does not apply to OTOCs at finite temperature can be found in \cite{lucas1903}.

\section{Memory Matrix Formalism}\label{app:MBL}
Here we review the memory matrix formalism (appropriate to our context).   Let $\mathfrak{p}$ be the projection operator onto all simple operators (at finite $T$), together with the current operator $|J)$.  (We include the components of $|J)$ which are not small, for technical reasons.)    As reviewed in \cite{lucasbook}, we may write a generalized conductivity matrix
 \begin{equation}
\hat{\sigma}_{AB}(\omega) =  \mathrm{i} (A| \mathfrak{p} (\mathcal{L}-\mathrm{i}\omega)^{-1} \mathfrak{p} |B) = \chi_{AC}\left(M + N - \mathrm{i}\omega \chi \right)^{-1}_{CD}\chi_{DB},  \label{eq:sigma}
\end{equation}
where \begin{subequations}\label{eq:3set}\begin{align}
 \chi_{AB}  &= \beta (A| \mathfrak{p} |B), \label{eq:chi} \\
N_{AB} &= \beta   (A|\mathfrak{p} \mathcal{L}\mathfrak{p} |B), \\
M_{AB}(\omega) &= \mathrm{i} \beta (A|\mathfrak{p}\mathcal{L}\mathfrak{q}  (\mathfrak{q}\mathcal{L}\mathfrak{q}-\mathrm{i}\omega)^{-1} \mathfrak{q}\mathcal{L}\mathfrak{p}   |B).  \label{eq:memmat}
\end{align}\end{subequations}
 $N_{AB}$ is antisymmetric and encodes the ``rotation" of small operators into one another.   All ``dissipation" of small operators into large operators is encoded in $M_{AB}$.     The electrical conductivity $\sigma$ is given by $\hat\sigma_{J_xJ_x}(0)$;  the matrix $\Gamma$ defined in the main text is simply $\Gamma = M(0)+N$.

\subsection{Eigenstates}
The main motivation for the remainder of the appendix is to explain why $\Gamma^{-1}$ is always antisymmetric in the $|J_x)$-direction whenever the conductivity vanishes.   We begin by evaluating the inner product (\ref{eq:innerproduct}) in a convenient basis of Hermitian operators.  Let $H|\alpha\rangle  = E_\alpha |\alpha\rangle$ correspond to an eigenstate/eigenvector of $H$.   Without loss of generality, we take $\langle \alpha|\beta\rangle = \mdelta_{\alpha\beta}$.   A suitable basis of Hermitian operators on $\mathcal{H}$ is \begin{subequations}\label{eq:eigenstates}\begin{align} 
|\alpha\alpha) &=  |\alpha\rangle\langle \alpha|, \\
 |\alpha\beta)_+ &=  \frac{|\alpha\rangle\langle \beta| -|\beta\rangle\langle \alpha| }{\sqrt{2}\mathrm{i}}, \\
  |\alpha\beta)_+ &=  \frac{|\alpha\rangle\langle \beta| +|\beta\rangle\langle \alpha| }{\sqrt{2}}.
\end{align}\end{subequations}
A straightforward calculation shows that this is an orthogonal set of basis vectors with \begin{equation}
\;_{\pm}(\alpha\beta|\alpha\beta)_{\pm} = \frac{1}{Z} \frac{\mathrm{e}^{-\beta E_\alpha} - \mathrm{e}^{-\beta E_\beta}}{E_\beta-E_\alpha}  \label{eq:appCchi}
\end{equation}
where $Z$ is the thermal partition function.  The Liouvillian $\mathcal{L}$ acts in a simple manner as well: \begin{subequations}\begin{align}
\mathcal{L}|\alpha\alpha) &= 0, \\
\mathcal{L}|\alpha\beta)_\pm &=  \mp (E_\alpha-E_\beta) |\alpha\beta)_\mp.
\end{align}\end{subequations}
We conclude that null vectors of $\mathcal{L}$ correspond to $|\alpha\alpha)$ and $|\alpha\beta)_\pm$ whenever $E_\alpha=E_\beta$.

\subsection{Conductivity}
In general, the conductivity vanishes not due to the divergence in any dissipative time scales, but due to the fact that acting with the current operator $J_x$ (which we denote as $J$ henceforth, for convenience) adds energy to the system.  More precisely, if $(J|\mathcal{O}) = 0$ whenever $\mathcal{L}|\mathcal{O}) = 0$, that means that $\langle \alpha|J|\beta\rangle = 0$ whenever $E_\alpha=E_\beta$.     It is then clear that $\sigma=0$:   we may write $\mathcal{L}$ as a block-diagonal matrix \begin{equation}
\mathcal{L} = \left(\begin{array}{cc} 0 &\ 0 \\ 0 &\ \mathcal{L}_0 \end{array}\right),
\end{equation}
where $\mathcal{L}_0$ is invertible, the top row/column corresponds to basis vectors which are null vectors of $\mathcal{L}$, and the bottom row/column corresponds to non-null vectors.    Using (\ref{eq:appCchi}), we see that the operator inner product is block diagonal. If $(J|\mathcal{O}) = 0$ whenever $\mathcal{L}|\mathcal{O}) = 0$, \begin{equation}
|J) = \left(\begin{array}{c} 0 \\ |J_0) \end{array}\right).
\end{equation}
Using (\ref{eq:sigma}), we conclude that \begin{equation}
\sigma = \lim_{\omega \rightarrow 0} \frac{1}{\mathrm{i}T}(J_0|(\mathcal{L}_0-\mathrm{i}\omega)^{-1}|J_0)= \frac{1}{\mathrm{i}T}(J_0|\mathcal{L}_0^{-1}|J_0)  = 0.
\end{equation}
In the last step, we used the fact that the inverse of an antisymmetric invertible matrix is antisymmetric.

We can also understand $\sigma=0$ from the memory matrix perspective.    Assuming that $\chi_{JJ}$ is not vanishing, which is guaranteed so long as (\ref{eq:innerproduct}) is a non-singular inner product,  then $\sigma=0$ implies $(M+N)^{-1}_{JJ}=0$.   (Without loss of generality we may consider an orthogonal basis of small operators where $\chi_{AB}$ is diagonal.)    If $(M+N)^{-1}_{JJ}=0$, then since $M+N$ is invertible, there exists an $X$ such that $(M+N)_{JX} \ne 0$.   Without loss of generality, we may freely rotate the orthogonal basis of operators to ensure $X$ is unique.   By construction of $N_{AB}$, this unique operator $X$ has overlap with $\dot{J} = \mathrm{i}[H,J]$.    Then using block matrix inversion identities we obtain $0=(M+N)^{-1}_{JJ} \propto (M+N)_{XX}=M_{XX}$.    The conductivity vanishes because $J$ rotates into an operator which has a vanishing decay rate.

\subsection{(Many-Body) Localized Phases}
We now show that in the MBL phase, $(J|\mathcal{O}) = 0$ whenever $\mathcal{L}|\mathcal{O}) = 0$.   The physical assumptions necessary to show this are (\emph{i}) in every eigenstate $\langle \alpha|J|\alpha\rangle = 0$ (which is plausible as these states do not transport charge), and (\emph{ii}) eigenstates are robust against small local perturbations \cite{serbyn}.   To be more precise in our second assumption, let $V$ be a sum of spatially local  operators, and let $|\alpha\rangle_\lambda $ denote a many-body eigenstate of $H+\lambda V$.  We then assume that  $\lVert \partial_\lambda |\alpha\rangle_\lambda \rVert < \infty$.     Thus there exist eigenstates $|\alpha\rangle_\lambda$, which vary continuously in $\lambda$, and whose eigenvalues $E_\alpha(\lambda)$ are continuous.   Applying $\partial_\lambda$ to $\langle \alpha|H|\beta\rangle =  E_\alpha \mdelta_{\alpha\beta}$ and $\langle \alpha|\beta\rangle = \mdelta_{\alpha\beta}$, and subsequently sending $\lambda \rightarrow 0$, we obtain \begin{equation}
\;_0 \langle \alpha|V|\beta \rangle_0 = (E_\beta(0)-E_\alpha(0))_0\langle \alpha| \partial_\lambda |\beta\rangle_0 \;\;\; (\alpha \ne \beta).
\end{equation}
Letting $V=J$, the current operator,  we conclude that $\langle \alpha |J|\beta\rangle \propto (E_\alpha-E_\beta)$.  Thus, as advertised, $|J)$ has no overlap with any null vector of $\mathcal{L}$.

%Next, we show that $\mathcal{L}_0^{-1}|J)$ has overlap with a simple vector, for sufficiently large (but $N$-independent) $R$.   To do so, simply observe that 

The arguments above also apply to the Anderson localized (non-interacting) insulator, with $\alpha$ and $\beta$ now single-particle eigenstates.

\end{appendix}

\bibliographystyle{unsrt}
\addcontentsline{toc}{section}{References}
\bibliography{localizationtransportbib}

\end{document}